\documentclass[12pt]{article}
\usepackage{epsf}
\title{ {\bf
The $b\rightarrow s g g$ decay in the two and three Higgs doublet models
with CP violating effects}}
\author{\vspace{1cm}\\
{\bf A. Goksu}
        \thanks{E-mail address:
        agoksu@metu.edu.tr}\,\,\, ,
{\bf E. O. Iltan}
        \thanks{E-mail address:
        eiltan@heraklit.physics.metu.edu.tr}\,\,\,  and 
{\bf L. Solmaz}
        \thanks{E-mail address:
        b107181@orca.cc.metu.edu.tr}
 \\
        Physics Department, Middle East Technical University \\
        Ankara, Turkey\\}

\date{}

\begin{document}
\setlength{\baselineskip}{24pt}
\maketitle
\setlength{\baselineskip}{7mm}
\begin{abstract}
We study the decay width and $CP$-asymmetry of the inclusive process 
$b\rightarrow s g g$ (g denotes gluon) in the three and two Higgs doublet  
models with complex Yukawa couplings. We analyse the dependencies of the 
differential decay width and $CP$-asymmetry to the $s$- quark energy 
$E_s$ and $CP$ violating parameter $\theta$. We observe that there exist 
a considerable enhancement in the decay width and $CP$ asymmetry is at the 
order of $10^{-2}$. Further, it is possible to predict the sign of 
$C_7^{eff}$ using the $CP$ asymmetry.
\end{abstract} 
\thispagestyle{empty}
\newpage
\setcounter{page}{1}
\section{Introduction}
Rare B-meson decays, induced by flavor changing neutral currents (FCNC), 
involve at the loop level in the standard model (SM), therefore they are 
phenomenologically rich. The measurements of the branching ratio ($Br$), 
$CP$ asymmetry ($A_{CP}$), forward-backward asymmetry, polarization effects, 
etc., provide restrictions on the SM parameters, such as the 
Cabibbo-Kobayashi-Maskawa (CKM) matrix elements, leptonic decay constants, 
etc. In addition, the possibility of replacing the SM particles by
non-standard ones results in the sensitivity of these decays beyond the SM,
like multi-Higgs doublet models, minimal supersymmetric extension of the SM
(MSSM) \cite{Hewett}, etc. The experimental effort at SLAC, KEK 
B-factories, HERA-B and possible future accelerators \cite{Bar,Ellis1} 
stimulate the theoretical studies on these rare decays. 

Among B-meson decay modes, inclusive $b\rightarrow s g$ decay becomes 
attractive since it is theoretically clean and affected by loop 
contributions due to new physics beyond the SM.
In the literature, there are various theoretical calculations on the $Br$ 
of this decay. In the SM, the Branching ratio of $b\rightarrow s g$ 
decay was calculated as $Br(b\rightarrow s g) \sim 0.2 \%$ for on-shell 
gluon \cite{Gao}. However, to decrease the averaged cham multiplicity  
$\eta_c$ \cite{Grizad} and to increase kaon yields \cite{Kagan} the 
enhancement of $Br(b\rightarrow s g)$ is helpful. The possibilities for 
the enhancement is the addition of the QCD corrections and non-standard 
effects coming from the new physics. In \cite{Chao1,Kagan2}, 
$Br\,(b\rightarrow s g)$ was calculated in the 2HDM  (Model I and II) 
for $m_{H^{\pm}}\sim 200\, GeV$ and $tan\,\beta\sim 5$ and it was found 
that there was an enhancement less than one order of magnitude. Further, 
this decay was studied in the supersymmetric models \cite{Berto} and in 
the framework of model III 2HDM \cite{Zhen}. In the model III, the 
enhancement was found at least one order of magnitude larger compared to 
the SM one and it was observed that there was no contradiction with the 
CLEO data \cite{CLEO1}
\begin{eqnarray} 
Br\,(b\rightarrow s g^*) < 6.8\,\% 
\label{Br1}
\end{eqnarray}   
for light-like gluon case. 
Recently, $O(\alpha_s)$ virtiual corrections and additional $O(\alpha_s)$ 
bremstrahlung effects to the decay width of 
$b\rightarrow s g$ was calculated in the SM \cite{GreubPatr} and the 
enhancement in the $Br$ was obtained as more than a factor of two larger of 
the previous SM results. 

The inclusive process $b\rightarrow s g g$ is another decay which has 
$Br$ at the same order as $Br\,(b\rightarrow s g)$ according to the studies 
in the literature  \cite{Hou1,Simma,Zhen2}. This process becomes not only 
from the chain decay  $b\rightarrow s g^*$ followed by 
$g^*\rightarrow g g$ but also from the emission of on-shell gluons from the 
quark lines to obey gauge invariance. In \cite{Simma}, the complete 
calculation was done in the collinear and infrared singularity free region, 
in the SM and $Br$ ratio was found at the order of magnitude $10^{-3}$. 
In \cite{Zhen,Zhen2} the additional contribution of gluon penguins in the 
model III was estimated as negligible. Recently $b\rightarrow s g g$ was
studied in the model III with real Yukawa couplings \cite{erilson} and 
a considerable enhancement was observed for the $Br$ of the process even          
2 orders of magnitude larger compared to the SM case.

In this work, we study the decay width $\Gamma$ and the $CP$ asymmetry 
$A_{CP}$ of $b\rightarrow s g g$ decay in the model III and the $3HDM(O_2)$. 
The reason to study the $b\rightarrow s g g$ process is the possible 
considerable enhancement of $\Gamma$ compared to the one in the SM and 
the measurable $A_{CP}$ in the framework of the models underconsideration. 
In our theoretical calculations  we choose the collinear and infrared 
singularity free kinematical region, following the procedure given in 
\cite{Simma}. Here we take the source of $CP$ violation as the complex 
Yukawa couplings appearing in both models.   

The paper is organized as follows:
In Section 2, we give a brief summary of the $3HDM(O_2)$ and  present the
calculation  of the decay width of the inclusive  
$b\rightarrow s g g $ decay in the framework of the $3HDM(O_2)$ and the 
model III. Further we calculate the (differential) $CP$-asymmetry 
($A_{CP}(E_s)$) $A_{CP}$ of the process. Section 3 is devoted to discussion 
and our conclusions. In Appendixes, we give the form factors appearing 
in the matrix element of the decay underconsideration and summarize the 
theoretical results for the $3HDM(O_2)$.
\section{The inclusive process $b\rightarrow s g g$ in the framework of the 
model III and $3HDM(O_2)$} 
The general Yukawa interaction in $3HDM$ is 
\begin{eqnarray}
{\cal{L}}_{Y}&=&\eta^{U}_{ij} \bar{Q}_{i L} \tilde{\phi_{1}} U_{j R}+
\eta^{D}_{ij} \bar{Q}_{i L} \phi_{1} D_{j R}+
\xi^{U}_{ij} \bar{Q}_{i L} \tilde{\phi_{2}} U_{j R}+
\xi^{D}_{ij} \bar{Q}_{i L} \phi_{2} D_{j R} \nonumber \\
&+&
\rho^{U}_{ij} \bar{Q}_{i L} \tilde{\phi_{3}} U_{j R}+
\rho^{D}_{ij} \bar{Q}_{i L} \phi_{3} D_{j R}
 + h.c. \,\,\, ,
\label{lagrangian}
\end{eqnarray}
where $L$ and $R$ denote chiral projections $L(R)=1/2(1\mp \gamma_5)$,
$\phi_{i}$ for $i=1,2,3$, are three scalar doublets and  
$\eta^{U,D}_{ij}$, $\xi^{U,D}_{ij}$, $\rho^{U,D}_{ij}$ are
the Yukawa matrices having complex entries, in general. With the 
choice of scalar Higgs doublets  
\begin{eqnarray}
\phi_{1}=\frac{1}{\sqrt{2}}\left[\left(\begin{array}{c c} 
0\\v+H^{0}\end{array}\right)\; + \left(\begin{array}{c c} 
\sqrt{2} \chi^{+}\\ i \chi^{0}\end{array}\right) \right]\, , 
\nonumber \\ \\
\phi_{2}=\frac{1}{\sqrt{2}}\left(\begin{array}{c c} 
\sqrt{2} H^{+}\\ H^1+i H^2 \end{array}\right) \,\, ,\,\, 
\phi_{3}=\frac{1}{\sqrt{2}}\left(\begin{array}{c c} 
\sqrt{2} F^{+}\\ H^3+i H^4 \end{array}\right) \,\, ,\nonumber
\label{choice}
\end{eqnarray}
and the vacuum expectation values,  
\begin{eqnarray}
<\phi_{1}>=\frac{1}{\sqrt{2}}\left(\begin{array}{c c} 
0\\v\end{array}\right) \,  \, ; 
<\phi_{2}>=0 \,\, ; <\phi_{3}>=0\,\, , 
\label{cb}
\end{eqnarray}
the SM particles carried by the first doublet and the information about 
the new physics by the others. The Yukawa interaction 
\begin{eqnarray}
{\cal{L}}_{Y,FC}=
\xi^{U}_{ij} \bar{Q}_{i L} \tilde{\phi_{2}} U_{j R}+
\xi^{D}_{ij} \bar{Q}_{i L} \phi_{2} D_{j R}
+\rho^{U}_{ij} \bar{Q}_{i L} \tilde{\phi_{3}} U_{j R}+
\rho^{D}_{ij} \bar{Q}_{i L} \phi_{3} D_{j R} + h.c. \,\, .
\label{lagrangianFC}
\end{eqnarray}
describes the Flavor Changing (FC) one beyond the SM.
Here, the couplings  $\xi^{U,D}$ and $\rho^{U,D}$ for the charged FC 
interactions are 
\begin{eqnarray}
\xi^{U}_{ch}&=& \xi_{N} \,\, V_{CKM} \nonumber \,\, ,\\
\xi^{D}_{ch}&=& V_{CKM} \,\, \xi_{N}  \nonumber \,\, , \\
\rho^{U}_{ch}&=& \rho_{N} \,\, V_{CKM} \nonumber \,\, ,\\
\rho^{D}_{ch}&=& V_{CKM} \,\, \rho_{N} \,\, ,
\label{ksi1} 
\end{eqnarray}
and
\begin{eqnarray}
\xi^{U,D}_{N}&=&(V_L^{U,D})^{-1} \xi^{U,D}\, V_R^{U,D}\,\, , \nonumber \\
\rho^{U,D}_{N}&=&(V_L^{U,D})^{-1} \rho^{U,D}\, V_R^{U,D}\,\, ,
\label{ksineut}
\end{eqnarray}
where the index "$N$" in $\xi^{U,D}_{N}$ denotes the word "neutral". 
Note that the Yukawa interaction for the model III can be obtained by 
taking into account only two doublets $\phi_1$, $\phi_2$ and Yukawa 
couplings $\eta^{U}_{ij}, \xi^{U}_{ij}$.

The decay amplitude of the process $b\rightarrow s g g$ is given by 
\begin{eqnarray}
M(b\rightarrow s gg)=i \frac{\alpha_s\, G_F}{\sqrt{2}\pi}
\epsilon_a^{\mu}(k_1)\epsilon_b^{\nu}(k_2) \bar{s}(p')T^{a\,b}_{\mu\nu}\,
b(p)\,\, ,
\label{Amp1}
\end{eqnarray}
where $\epsilon_a^{\mu}(k)$ are polarization vectors of the gluons with
color $a$ and momentum $k$, 
\begin{eqnarray}
T^{a\,b}_{\mu\nu}=T_{\mu\nu}\frac{\lambda^b}{2}\,\frac{\lambda^a}{2}+
T^{E}_{\mu\nu}\,\frac{\lambda^a}{2}\,\frac{\lambda^b}{2}\,\, .
\label{Tfunc}
\end{eqnarray}
Here $\frac{\lambda^a}{2}$ are the Gell-Mann matrices and $T^{E}_{\mu\nu}$ 
can be obtained by the replacements $k_1\leftrightarrow k_2$, 
$\mu\leftrightarrow \nu$ in the function $T_{\mu\nu}$. 

Since the process occurs at least at one-loop level in the SM, the function 
$T_{\mu\nu}$ have contributions coming from light and heavy internal quarks, 
namely, $u,c,t$. Internal quarks $d,s,b$ can also give
contribution to the process beyond the SM. In the case of heavy internal 
quark, $t$-quark, the terms $k^2_{external}/m_i^2\, 
(m^2_W,\, m^2_{H^{\pm}},\,m^2_{F^{\pm}})$ are neglected. However, for light 
internal quarks, $k^2_{external}/m_i^2$ terms can give considerable 
contribution.  This forces us to parametrize the function $T_{\mu\nu}$ as   
\begin{eqnarray}
T_{\mu\nu}=T^{heavy}_{\mu\nu}+T^{light}_{\mu\nu}\,\, ,
\label{Tfunctot1}
\end{eqnarray}
(for the explicit forms of $T^{heavy}_{\mu\nu}$ and 
$T^{light}_{\mu\nu}$ see Appendix A).
On the other hand, $T^{a\,b}_{\mu\nu}$ can be divided into color symmetric 
and antisymmetric parts as \cite{Simma} 
\begin{eqnarray}
T_{\mu\nu}^{a\,b}=T_{\mu\nu}^{+} \{ \frac {\lambda^{b}}{2}, 
\frac {\lambda^{a}}{2} \}+ T_{\mu\nu}^{-}[\frac {\lambda^{b}}{2},
\frac {\lambda^{a}}{2}]\,\, ,
\label{Tfunc2}
\end{eqnarray}  
with
\begin{eqnarray}
T_{\mu\nu}^{+}&=&\frac{1}{2}(T_{\mu\nu}+T_{\mu\nu}^{E}) 
\nonumber\,\, ,\\
T_{\mu\nu}^{-}&=&\frac{1}{2}(T_{\mu\nu}-T_{\mu\nu}^{E})\,\, . 
\label{Tfunc3}
\end{eqnarray}
Finally, using the expression 
\begin{eqnarray}
\Gamma^{Sym\,(Asym)} \sim Tr(C_+(-) T_{\mu\nu}^{+\,(-)}\, (\not\!{p}+m_b))\,
\bar{T}_{\mu '\nu '}^{+\,(-)}\,\not\!{p'})\,P^{\mu\mu '}\,P^{\nu\nu '} 
\,\, , 
\end{eqnarray}
with the color factors 
$C_+=\frac{(N_c^2-1)(N_c^2-2)}{2\, N_c}$ and $C_-=\frac{N_c\,(N_c^2-1)}{2}$\, 
and the polarization sum of the on-shell gluons 
\begin{eqnarray}
P^{\mu\mu '}=-g^{\mu\mu '}+\frac{ k_1^{\mu}\,k_2^{\mu '}+
k_2^{\mu}\,k_1^{\mu '}}{k_1.k_2} \nonumber \,\, ,
\end{eqnarray}
we get the differential decay width of the process
\begin{eqnarray}
\frac{d^2\,\Gamma}{dE_s\,dE_1}=\frac{1}{2\pi^3}\frac{1}{8\, m_b}
|\bar{M}|^2\,\, .
\label{diffCS}
\end{eqnarray}
Here  $E_s$ is the $s$-quark energy and $E_1$ is the energy of gluon with
polarization $\epsilon_{\mu}^{a}(k_1)$. $\bar{M}$ is the average 
decay amplitude, $\bar{M}=\frac{1}{2\,J+1}\,\frac{1}{N_c}\, M$, and 
$J=\frac{1}{2}$, $N_c=3$.

In the expressions, the symmetric and antisymmetric parts do not mix
each other. Further, the decay width can be divided into three sectors 
(see Appendix A):left ($\Gamma^L$), right ($\Gamma^R$) and left-right 
($\Gamma^{LR}$). Left one is created by the 
nonvanishing $k_{external}^2/m_{light}^2$ terms, however right sector 
contains the forms factors with parameters $m_i^2/m_W^2$ and $m_i^2/m_H^2$ 
where $i=u,c,t$ and  $H$ is one of the Higgs bosons. Left-right sector 
contains mixed terms.

Now we are ready to calculate the $CP$-violating asymmetry $A_{CP}$ of the
process $b\rightarrow s g g$. In the model III and $3HDM(O_2)$, the complex
Yukawa couplings are possible sources for $CP$ violation. Our procedure is 
to neglect neutral boson effects  and all Yukawa 
couplings except $\bar{\xi}^U_{N,tt}$ and $\bar{\xi}^D_{N,bb}$ 	
($\bar{\epsilon}^U_{N,tt}$ and $\bar{\epsilon}^D_{N,bb}$) (see Discussion) 
in the model III ($3HDM(O_2)$) (see Appendix B for the definitions of 
$\bar{\epsilon}^U_{N,tt}$ and $\bar{\epsilon}^D_{N,bb}$). Therefore, in 
the model III ($3HDM(O_2)$), only the combination 
$\bar{\xi}^U_{N,tt} \bar{\xi}^D_{N,bb}$ 
($\bar{\epsilon}^U_{N,tt} \bar{\epsilon}^D_{N,bb}$) is responsible for 
$A_{CP}$. 
Using the parametrization
\begin{eqnarray}
\lambda_{\theta}= \left\{
\begin{array}{ll}
\frac{1}{m_t\, m_b}\, \bar{\xi}^{U}_{N,tt}\, 
\bar{\xi}^{D}_{N, bb}\, e^{i\,\theta}  &\mbox{(model \,\, III)}
\nonumber \,\, , \\ \\ 
\frac{1}{m_t\, m_b}\, \bar{\epsilon}^{U}_{N,tt}\,
\bar{\epsilon}^{D}_{N, bb}\,
(cos^2\, \theta+ i sin^2 \, \theta) \,\, &(3HDM(O_2)) \,\, ,
\end{array}
\right.
\label{lambdatheta}
\end{eqnarray}
\\
and the definition of differential $CP$ asymmetry 
$A_{CP}(E_s)$ 
\begin{eqnarray}
A_{CP}(E_s)=\frac{\frac{d^2\,\Gamma}{dE_s\,dE_1}(b\rightarrow s g g)-
\frac{d^2\,\Gamma}{dE_s\,dE_1} (\bar{b}\rightarrow \bar{s} g g)}
{\frac{d^2\,\Gamma}{dE_s\,dE_1}(b\rightarrow s g g)+
\frac{d^2\,\Gamma}{dE_s\,dE_1} (\bar{b}\rightarrow \bar{s} g g)},\,\,
\label{ACPEs}
\end{eqnarray}
we get 
\begin{eqnarray}
A_{CP}(E_s)=2 Im(\lambda_{\theta})\, G_2\,(y_t)\, 
\frac{(Im(\Delta\, F_1-\Delta\, i_2))\, 
\Omega}
{\Lambda}
\label{ACPEs1}
\end{eqnarray}
where
\begin{eqnarray}
\Omega &=& 18\, E_1 \,m_b \,
(2 \,E1-m_b) (-2 \,E_s + m_b)\,( (2\, E_s-m_b)\, m_b+ 2\, E_1\, (2\, E_s + 
m_b)) 
\nonumber \,\, , \\
\Lambda &=& -2\, |\tilde{F_2}|^2\, m_b\, (2496\, E_1^5 + 192\, E_1^4\, 
(20\, E_s-23\, m_b) + m_b^3\, (-28\, E_1^2 + 44\, E_s\, m_b-15\, m_b^2)
\nonumber \\
&+&  2\,E_1\, m_b^2\, (172\, E_s^2-208\, E_s\, m_b + 69\, m_b^2)-
4\, E_1^2\, m_b\, (316 \,E_s^2-562\, E_s\, m_b + 213\, m_b^2)
\nonumber \\ &+& 
8\, E_1^3\, (204\, E_s^2 - 638\, E_s\, m_b + 357 \,m_b^2) ) \nonumber \\
&+& 2\, Re (\tilde{F_2})\, Re(\Delta\, F_1-\Delta\, i_2)\,\Omega + 
8\, E_1^2\, (2\, E_1- m_b)\,(-2\, E_s + m_b)
(7\, |\Delta\, i_5|^2\, 
E_s\, m_b + 9\, |\Delta\, i_2|^2\, 
\nonumber \\ & & 
(8\, E_1^2 + 8\, E_1\, (E_s - m_b) + m_b\, (-3\, E_s + 2\, m_b))
\nonumber \\
&+& 9\, |\Delta\, F_1|^2\, (8 \,E_1^2+ 8\, E_1\, (E_s-m_b)+ m_b\, 
(-3\, E_s + 2\, m_b))
\nonumber \\
&-& 18\, Re(\Delta\, F_1^* \, \Delta\, i_2)\, (8\, (E_1^2 + E_1\, E_s - 
E_1\, m_b) - 3\, E_s\, m_b + 2\, m_b^2 ) \,\, . 
\label{ACPEs2}
\end{eqnarray}
Here $\theta$ is the $CP$ violating parameter which is restricted by the
experimental upper limit of the neutron electric dipole moment eq. 
(\ref{neutrdip}) and $\tilde{F_2}=F_2^{3HDM}-F_2^{SM}(0)$, 
$\Delta\, F_1$, $\Delta\, i_2$, and $\Delta\, i_5$ are the Wilson
coefficients (eqs.(\ref{Ffunc}) and 
(\ref{F1i2i5})). 
For the calculation of the $CP$ asymmetry $A_{CP}$ 
\begin{eqnarray}
A_{CP}=\frac{\Gamma(b\rightarrow s g g)-
\Gamma(\bar{b}\rightarrow \bar{s} g g)}
{\Gamma(b\rightarrow s g g)+
\Gamma(\bar{b}\rightarrow \bar{s} g g)},\,\,
\label{ACP}
\end{eqnarray}
the integration over $E_1$ and $E_s$ should be done. However there are 
collinear divergences at the boundary of the kinematical region. 
To overcome these divercences we follow the procedure given in \cite{Simma}, 
namely taking a cutoff $c$ in the integration over phase space as: 
\begin{eqnarray}
 \frac{m_b}{2}-E_s \leq \, E_1\,\leq
\frac{m_b}{2} (1-c) \,\, ,
\label{cutoff}
\end{eqnarray}
and 
\begin{eqnarray}
c \frac{m_b}{2}\leq \, E_s\,\leq
\frac{m_b}{2} (1-c) \,\, ,
\label{cutoff2}
\end{eqnarray}
with $c=0.1$.
Note that left-right sector gives small contribution to $\Gamma$, however
this part is responsible for the $A_{CP}$. Further $A_{CP}$ contains only 
antisymmetric sector.  

\section{Discussion}
In the general 3HDM model, there are many free parameters, such as 
masses of charged and neutral Higgs bosons, complex Yukawa matrices, 
$\xi_{ij}^{U,D}$, $\rho_{ij}^{U,D}$ where $i,j$ are quark flavor indices. 
The additional global $O(2)$ symmetry in the Higgs flavor space   
connects the Yukawa matrices in the second and third doublet  
and also keeps  the masses of new charged (neutral) Higgs particles 
in the third doublet to be the same as the ones in the second doublet 
(Appendix B). Further, the Yukawa couplings, which are entries of Yukawa 
matrices, can be restricted using the experimental measurements, 
$\Delta F=2$ mixing, the $\rho$ parameter \cite{Soni} and the CLEO 
measurement \cite{CLEO2}, 
\begin{eqnarray}
Br (B\rightarrow X_s\gamma)= (3.15\pm 0.35\pm 0.32)\, 10^{-4} \,\, .
\label{br2}
\end{eqnarray}
In our calculations, we neglect all Yukawa couplings except 
$\bar{\xi}^{U}_{N,tt}$ ,$\bar{\xi}^{D}_{N,bb}$, 
$\bar{\rho}^{U}_{N,tt}$ and $\bar{\rho}^{D}_{N,bb}$ by respecting these 
measurements. The same restrictions are done in the model III case and only 
$\bar{\xi}^{U}_{N,tt}$ and $\bar{\xi}^{D}_{N,bb}$ are taken into account.  

In this section, we study the $s$ quark energy $E_s$ dependency of the 
differential decay width $\frac{d\,\Gamma}{d\,E_s}$, differential
$CP$-asymmetry $A_{CP}(E_s)$ and the parameter $sin\theta$ dependency of 
the decay width $\Gamma$, $CP$-asymmetry $A_{CP}$ for the inclusive decay 
$b\rightarrow s g g$ in the framework of the model III and $3HDM(O_2)$. 
In our analysis, we restrict the parameters  $\theta$,
$\bar{\epsilon}^{U}_{N,tt}$ and  $\bar{\epsilon}^{D}_{N bb}$ 
($\bar{\xi}^{U}_{N,tt}$ and $\bar{\xi}^{D}_{N bb}$ in the model III) 
using the constraint for $|C_7^{eff}|$, $0.257 \leq |C_7^{eff}| \leq 0.439$ 
where the upper and lower limits were calculated in \cite{alil1} following 
the procedure given in \cite{gudalil}. Here $C_7^{eff}$ is the effective 
magnetic dipole type Wilson coefficient for $b\rightarrow s\gamma$ 
vertex (see \cite{alil1}). The above restriction allows us to define a 
constraint region for the parameter $\bar{\epsilon}^{U}_{N,tt}$ 
($\bar{\xi}^{U}_{N,tt}$) in terms of $\bar{\epsilon}^{D}_{N,bb}$ 
($\bar{\xi}^{D}_{N,bb}$) and $\theta$ in the $3HDM(O_2)$ (the model \, III). 
Our numerical calculations based on this restriction and the constraint for 
the angle $\theta$, due to the experimental upper limit of neutron electric 
dipole moment, namely  
\begin{eqnarray}
d_n<10^{-25}\hbox{e$\cdot$cm}
\label{neutrdip}
\end{eqnarray}
which places an upper bound on the couplings with the expression in 
$3HDM(O_2)$ ($model \, III$): $\frac{1}{m_t m_b} 
(\bar{\epsilon}^{U}_{N,tt}\,\bar{\epsilon}^{* D}_{N,bb})sin^2\,
\theta < 1.0$ ($\frac{1}{m_t m_b} (\bar{\xi}^{U}_{N,tt}\,
\bar{\xi}^{* D}_{N,bb})sin\,\theta < 1.0$) for $m_{H^\pm}\approx 200$ GeV 
\cite{David}. 
Throughout these calculations, we take the charged Higgs mass
$m_{H^{\pm}}=400\, GeV$, and we use the input values given in Table 
(\ref{input}).  
\begin{table}[h]
        \begin{center}
        \begin{tabular}{|l|l|}
        \hline
        \multicolumn{1}{|c|}{Parameter} & 
                \multicolumn{1}{|c|}{Value}     \\
        \hline \hline
        $m_c$                   & $1.4$ (GeV) \\
        $m_b$                   & $4.8$ (GeV) \\           
        $\lambda_t$            & 0.04 \\
        $m_{t}$             & $175$ (GeV) \\
        $m_{W}$             & $80.26$ (GeV) \\
        $m_{Z}$             & $91.19$ (GeV) \\
        $\Lambda_{QCD}$             & $0.214$ (GeV) \\
        $\alpha_{s}(m_Z)$             & $0.117$  \\
        $c$                  & $0.1$  \\
        \hline
        \end{tabular}
        \end{center}
\caption{The values of the input parameters used in the numerical
          calculations.}
\label{input}
\end{table}

\begin{figure}[htb]
\vskip -3.0truein
\centering
\epsfxsize=6.8in
\leavevmode\epsffile{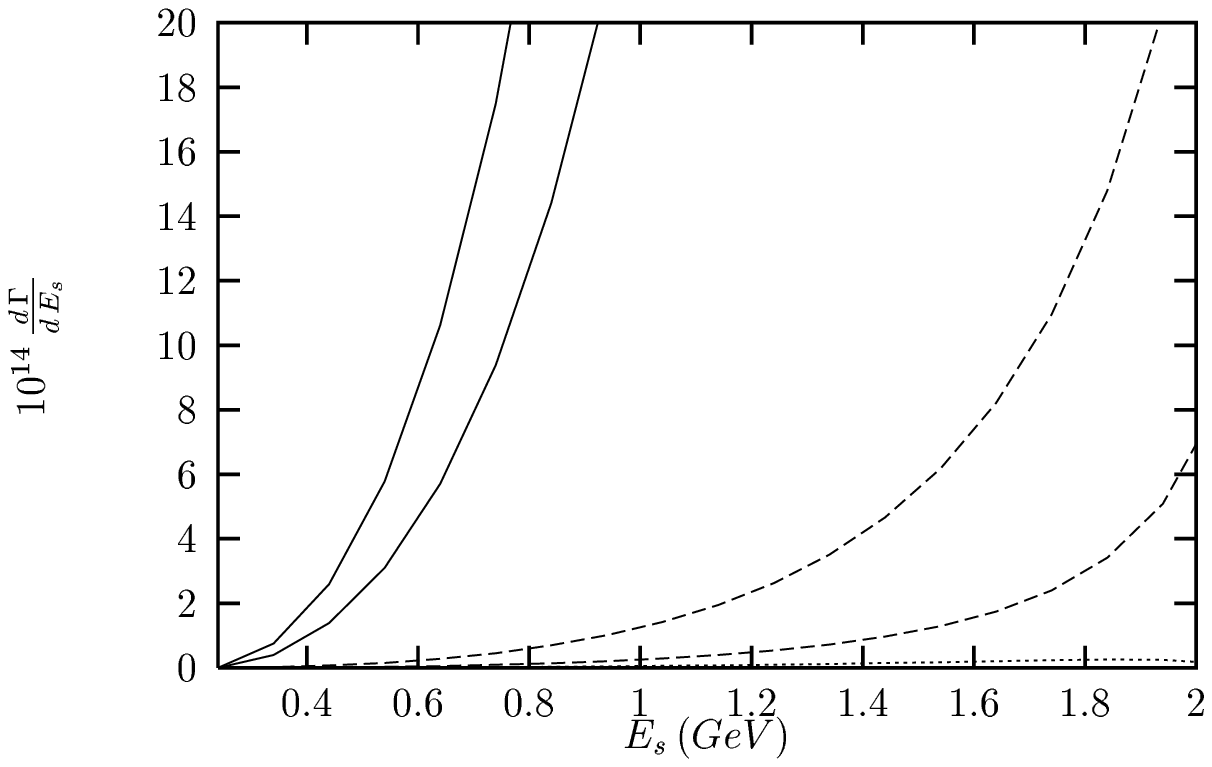}
\vskip -3.0truein
\caption[]{$\frac{d\,\Gamma}{d\,E_s}$ as a function of 
$E_s$  for fixed $\bar{\xi}_{N,bb}^{D}=40\, m_b$, $sin\theta=0.5$ and  
$|r_{tb}|=|\frac{\bar{\xi}_{N,tt}^{U}}{\bar{\xi}_{N,bb}^{D}}| <1$ in the 
model III. Here $\frac{d\,\Gamma}{d\,E_s}$  is restricted in the region 
bounded by solid (dashed) lines for $C_7^{eff} > 0$ ($C_7^{eff} < 0$). 
Dotted line represents the SM contribution.}
\label{dGammaEs}
\end{figure}

\begin{figure}[htb]
\vskip -3.0truein
\centering
\epsfxsize=6.8in
\leavevmode\epsffile{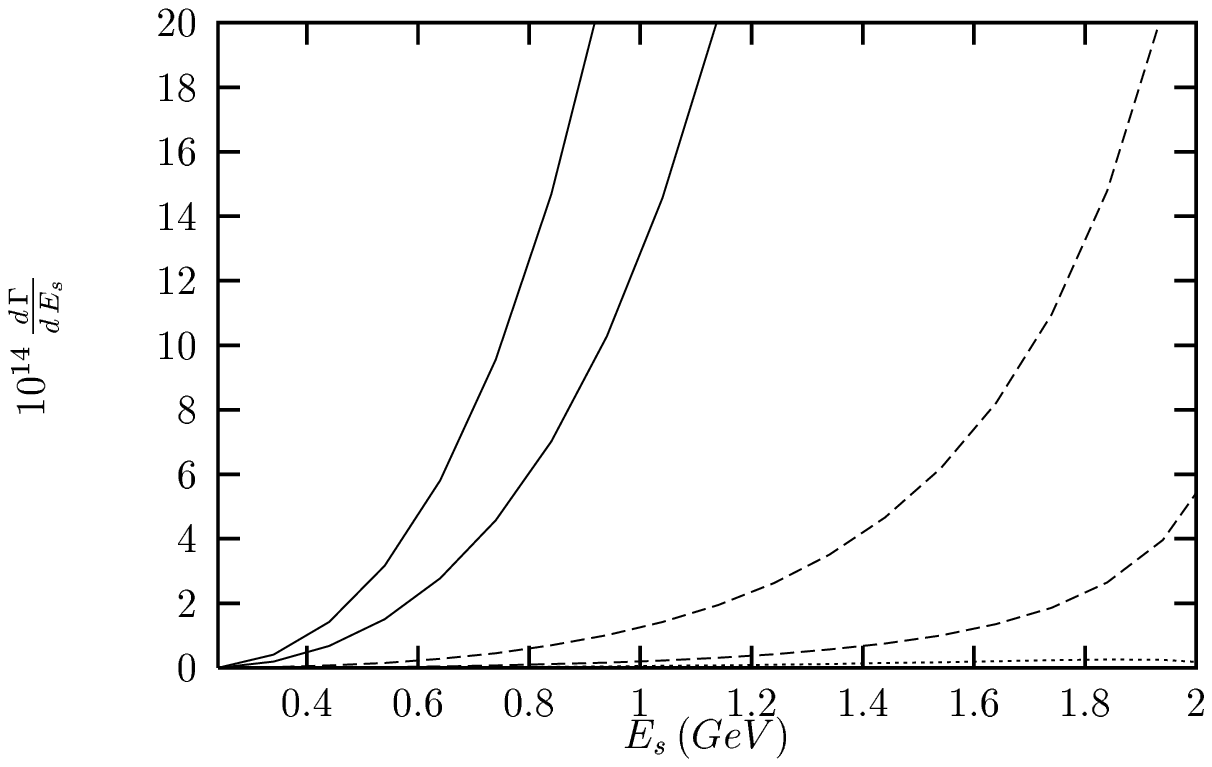}
\vskip -3.0truein
\caption[]{The same as Fig. \ref{dGammaEs} but for $3HDM(O_2)$.}
\label{dGamma3HEs}
\end{figure}

In  Fig.~\ref{dGammaEs} (\ref{dGamma3HEs}) we plot
$\frac{d\,\Gamma}{d\,E_s}$ 
with respect to the $s$ quark energy $E_s$, for $sin\theta=0.5$, 
$\bar{\xi}_{N,bb}^{D}=40\, m_b$ and 
$|r_{tb}|=|\frac{\bar{\xi}_{N,tt}^{U}}{\bar{\xi}_{N,bb}^{D}}| <1.$
$\frac{d\,\Gamma}{d\,E_s}$  is restricted in the region bounded by solid
(dashed) lines for $C_7^{eff} > 0$ ($C_7^{eff} < 0$). Dotted line represents 
the SM contribution. There is a large enhancement in the differential
decay width for $C_7^{eff} > 0$ and $C_7^{eff} < 0$ in both models. (see
\cite{erilson} for the model III with real Yukawa couplings). 
In the $3HDM(O_2)$, the enhancement is smaller and the restriction 
regions are broader compared to the ones in the model III.

\begin{figure}[htb]
\vskip -3.0truein
\centering
\epsfxsize=6.8in
\leavevmode\epsffile{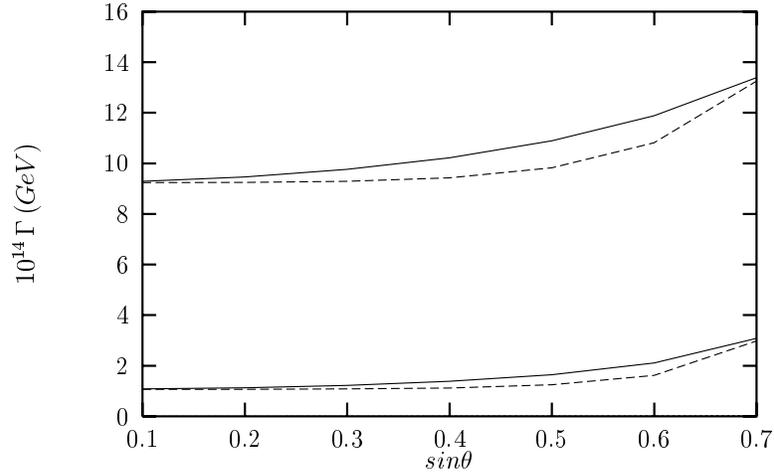}
\vskip -3.0truein
\caption[]{$\Gamma$ as a function of 
$sin\,\theta$ for $C_7^{eff} < 0$, $\bar{\xi}_{N,bb}^{D}=40\, m_b$, 
and $|r_{tb}|<1$. Here $\Gamma$  is restricted in the region bounded 
by solid (dashed) lines for the model III ($3HDM(O_2)$).} 
\label{Gammasin2H3H}
\end{figure}

Fig.~\ref{Gammasin2H3H} is devoted to the $sin\theta$ dependence of 
$\Gamma$ for $\bar{\xi}_{N,bb}^{D}=40\, m_b$ and $C_7^{eff} < 0$  in the 
region $|r_{tb}| <1$. Here $\Gamma$ in the model III $(3HDM(O_2))$ is 
restricted in between solid (dashed) lines. As shown in the figure, the 
decay width $\Gamma$ increases with increasing $sin\theta$. Further, the 
upper and lower bounds for $\Gamma$ in the model III exceed the ones in 
the $3HDM(O_2)$ especially for the intermediate values of the parameter 
$sin\theta$. Further $\Gamma$ can reach the value $10^{-3}\, GeV$ in both
models and this is a considerable enhancement compared to the SM one, which
is at the order of magnitude $10^{-5}\, GeV$ (see \cite{Simma,erilson}).   
 
\begin{figure}[htb]
\vskip -3.0truein
\centering
\epsfxsize=6.8in
\leavevmode\epsffile{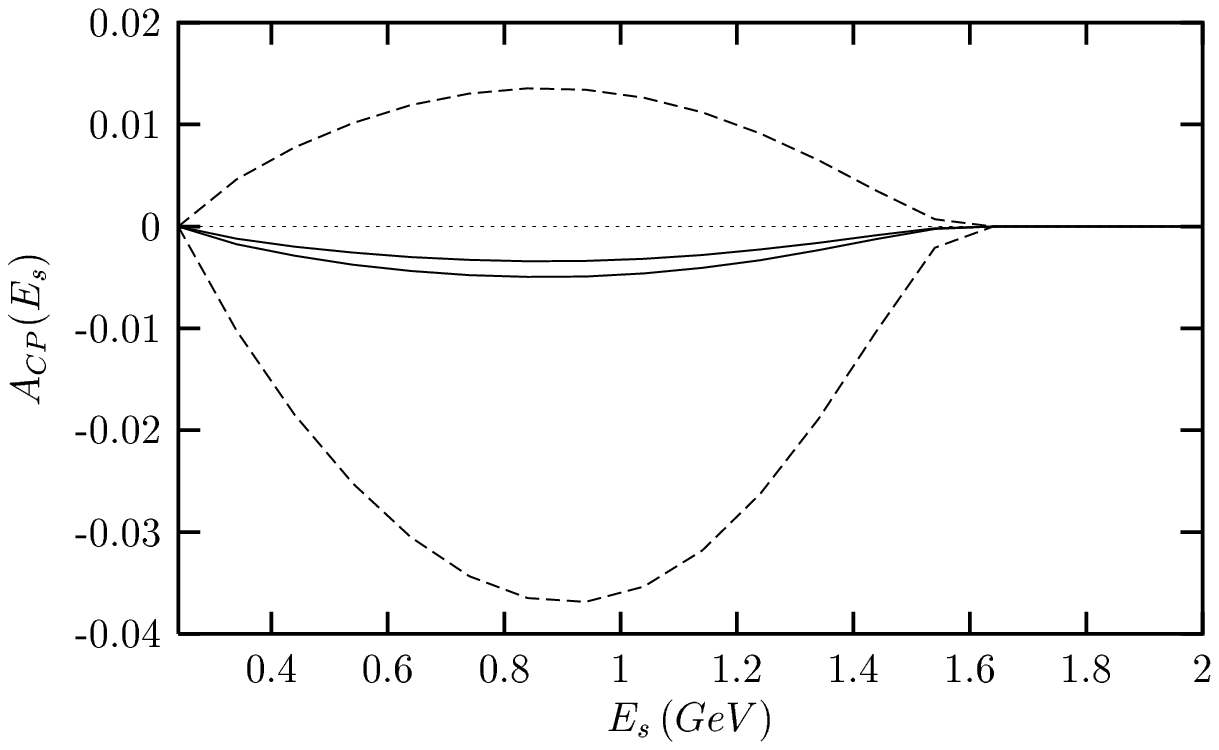}
\vskip -3.0truein
\caption[]{$A_{CP}(E_s)$ as a function of $E_s$  for fixed 
$\bar{\xi}_{N,bb}^{D}=40\, m_b$, $sin\theta=0.5$ and  
$|r_{tb}|<1$ in the model III. Here $A_{CP}(E_s)$ is restricted in the 
region bounded by solid (dashed) lines for $C_7^{eff} > 0$ 
($C_7^{eff} < 0$).}
\label{Cp2HEs}
\end{figure}

\begin{figure}[htb]
\vskip -3.0truein
\centering
\epsfxsize=6.8in
\leavevmode\epsffile{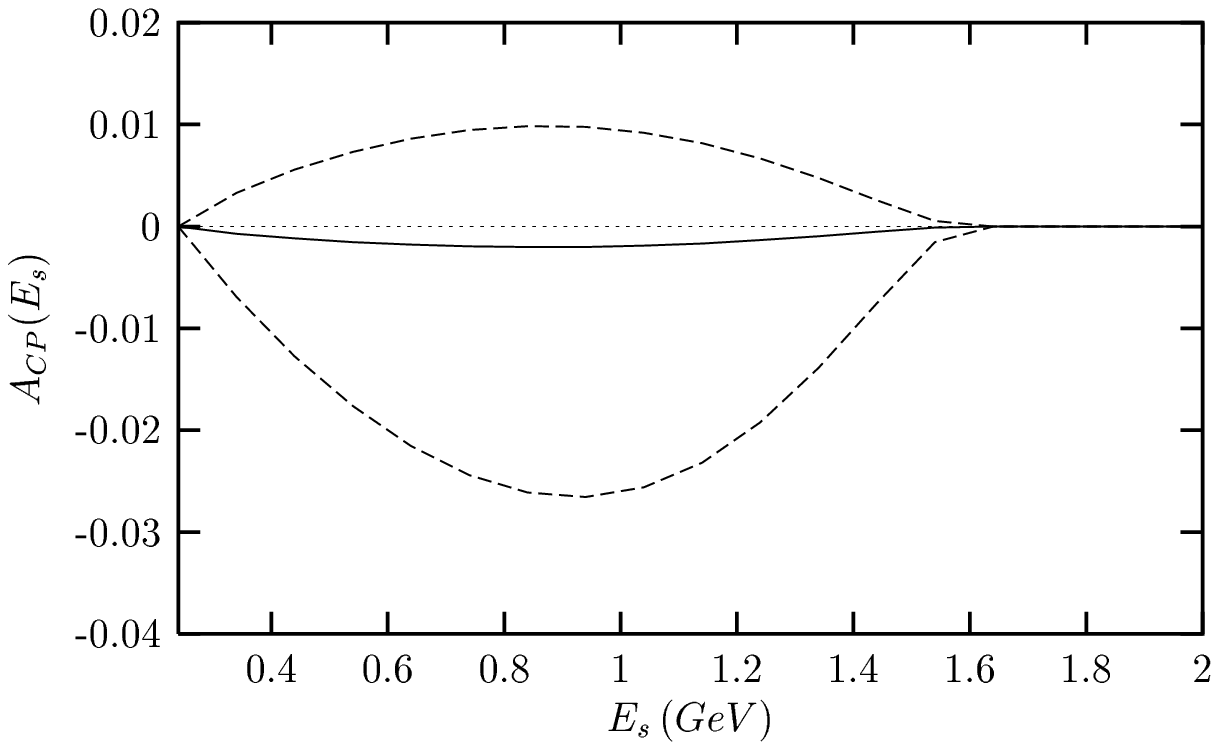}
\vskip -3.0truein
\caption[]{The same as Fig. \ref{Cp2HEs}, but for $3HDM(O_2)$.}
\label{Cp3HEs}
\end{figure}

Fig. \ref{Cp2HEs} (\ref{Cp3HEs}) shows the $E_s$ dependence of 
$A_{CP}(E_s)$ in the model III ($3HDM(O_2)$). Here solid (dashed) lines are
the boundaries of the allowed regions of $A_{CP}(E_s)$ for $C_7^{eff} > 0$
($C_7^{eff} < 0$). In model III, the restriction region for $A_{CP}(E_s)$ 
is narrow for $C_7^{eff} > 0$ and it has only negative values at the order 
of magnitude $10^{-3}$. However, for $C_7^{eff} < 0$, this region is broader 
and contains both negative and positive values. The possible values of 
$|A_{CP}(E_s)|$ reaches $\sim 4 \%$ for $0.8\, GeV \leq E_s \leq 1.0 \, GeV$. 
In the $3HDM(O_2)$, upper and lower boundaries of the allowed region for 
$A_{CP}(E_s)$ are almost coincides for $C_7^{eff} > 0$  and this region 
becomes narrower for $C_7^{eff} < 0$ compared to the one in the model III. 
In this model $|A_{CP}(E_s)|$ can be $\sim 2.5\,\%$ as a maximum value.

\begin{figure}[htb]
\vskip -3.0truein
\centering
\epsfxsize=6.8in
\leavevmode\epsffile{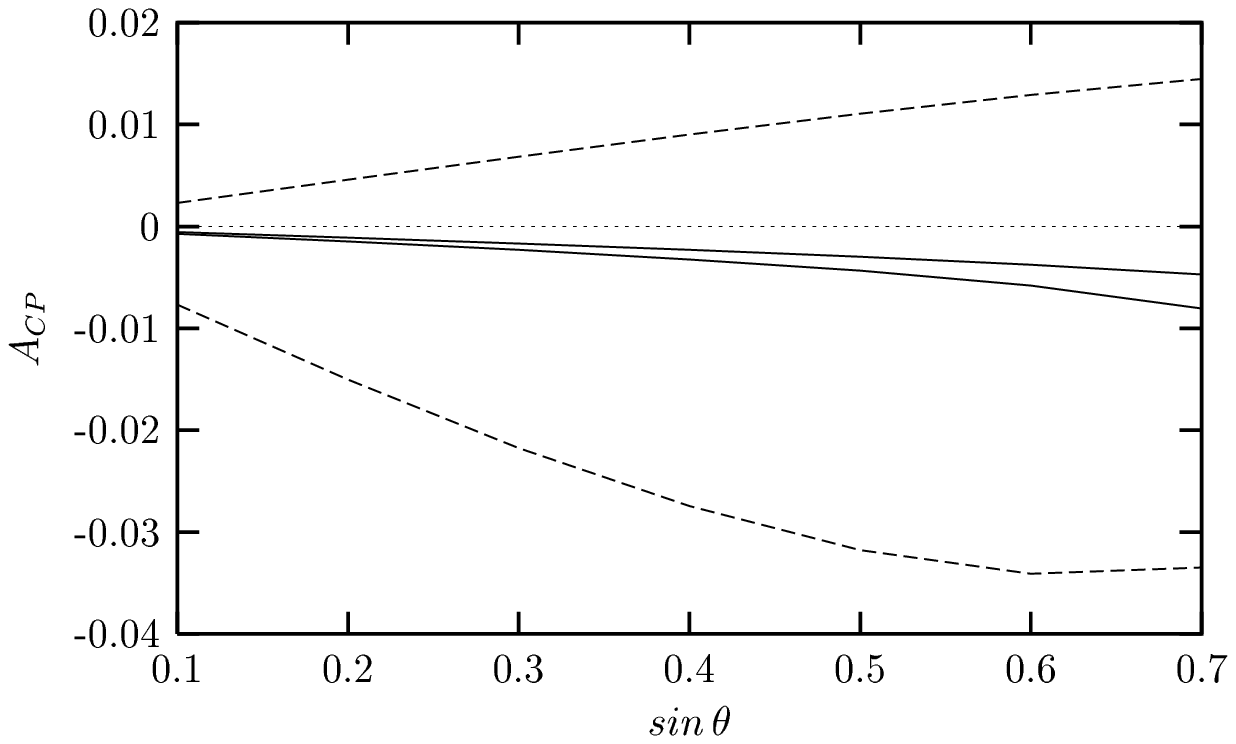}
\vskip -3.0truein
\caption[]{$A_{CP}$ as a function of $sin\,\theta$  for  
$\bar{\xi}_{N,bb}^{D}=40\, m_b$  and  
$|r_{tb}|<1$ in the model III. Here $A_{CP}$ is restricted in the 
region bounded by solid (dashed) lines for $C_7^{eff} > 0$ 
($C_7^{eff} < 0$)}.
\label{Cp2Hsin}
\end{figure}

\begin{figure}[htb]
\vskip -3.0truein
\centering
\epsfxsize=6.8in
\leavevmode\epsffile{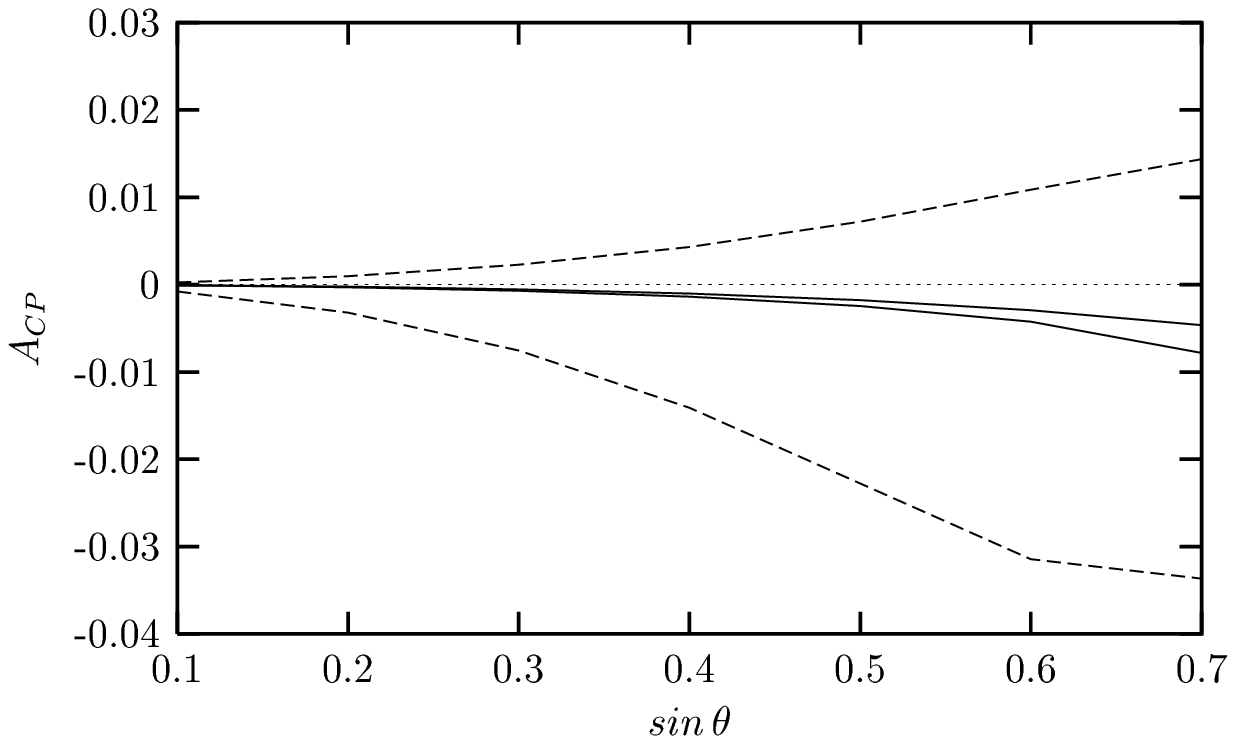}
\vskip -3.0truein
\caption[]{The same as Fig \ref{Cp2Hsin}, but for $3HDM(O_2)$.}
\label{Cp3Hsin}
\end{figure}

In  Fig.~\ref{Cp2Hsin} and Fig.~\ref{Cp3Hsin}, we represent 
the $sin\theta$ dependence of $A_{CP}$ in the model III and $3HDM(O_2)$.
$A_{CP}$ is  restricted in the narrow region bounded by solid lines 
for $C_7^{eff} > 0$ and it reaches $-0.8\,\%$ for $sin\theta=0.7$ in both 
models. All possible values of $A_{CP}$ are negative in this case. However, 
for $C_7^{eff} < 0$, allowed region becomes broader and $A_{CP}$ can take 
positive and negative values. For this case, $|A_{CP}|$ can reach $3.4\,\%$.  
Note that the restricted regions are broader in the model III compared to
the ones in $3HDM(O_2)$. 

As a conclusion, we get an enhancement in the decay width of the
process $b\rightarrow s g g $ in both models, model III and $3HDM(O_2)$. 
This enhancement is too large to respect the total decay width 
$\Gamma^{tot} (b\rightarrow s X)=3.50\pm 1.50\, 10^{-3}\, GeV$ 
for $C_7^{eff} > 0$. For $C_7^{eff} < 0$, $\Gamma$ can reach the values 
more than two orders of magnitude larger compared to the SM case.      
Further, we study $A_{CP}$ of the process $b\rightarrow s g g $ in both
models. In the SM, the only source for the CP-violation is the complex
Cabbibo-Kobayashi-Maskawa matrix elements and almost there is no 
violation  for this process. However in the model III and the 
$3HDM(O_2)$, the absolute value of $A_{CP}$ can reach to $3-4\,\%$, which 
is a measurable quantity. In addition, we observe that $C_7^{eff}$ is 
necessarily negative if $A_{CP}$ has positive values. Therefore the 
experimental study of the decay width $\Gamma$ and $A_{CP}$  of the process 
$b\rightarrow s g g $ can give important clues for the physics beyond 
the SM and also the sign of $C_7^{eff}$.
\newpage
{\bf \LARGE {Appendix}} \\
\begin{appendix}
\section{The form factors appearing in the $b\rightarrow s g g$ decay}
The function $T_{\mu\nu}$ can be divided into two parts:  
\begin{eqnarray}
T_{\mu\nu}=T^{heavy}_{\mu\nu}+T^{light}_{\mu\nu}\,\, \nonumber.
\end{eqnarray}
Here $T^{heavy}_{\mu\nu}$ is  the contribution due to the heavy internal 
quark and neglecting $s$-quark mass, it is given as 
\begin{eqnarray}
T^{heavy}_{\mu\nu}&=&
-i \, \lambda_t \, F^{3HDM}_2 \, \{ \big ( \frac{2 \,p'_{\nu}+
\gamma_{\nu} \not\!{k_2}}{2 \,p'.k_2}\,
\sigma_{\mu\alpha} k_1^{\alpha}+\sigma_{\nu\alpha} k_2^{\alpha}\,
\frac{2 p_{\mu}-\not\!{k_1} \gamma_{\mu}}{-2 p.k_1} \big )\nonumber \\ 
&+& 
\frac{1}{q^2} \big ( 2\, \sigma_{\alpha\beta} k_{1}^{\alpha}\, 
k_2^{\beta} g_{\mu\nu}+2\, \sigma_{\nu\alpha} k_{2\,\mu}\, q^{\alpha}-
2\, \sigma_{\mu\alpha} k_{1\,\nu}\, q^{\alpha}+ 
\sigma_{\mu\nu} q^2 \big ) \} \,m_b\, R\,\, .  
\label{Tfunch}
\end{eqnarray}
where $q$ is the momentum transfer, $q=k_1+k_2$, $\lambda_t$ is the CKM 
matrix combination $\lambda_t=V_{tb} V^{*}_{ts}$ and $F^{3HDM}_2$ is the 
form factor 
\begin{eqnarray}
F^{3HDM}_2=F^{SM}_2\,(x_t)+F^{Beyond}_2\,(y_t, y_t^{\prime}).
\label{Ffunc}
\end{eqnarray}
In eq. (\ref{Ffunc}), $F^{SM}_2\,(x_t)$ is the magnetic dipole form factor 
of $b\rightarrow s g^*$ vertex 
\begin{eqnarray}
F_2^{SM}(x_t)=\frac {-8+38 \,x_{t}-39 \,x^{2}_{t}+14\, x^{3}_{t}-5\,
x^{4}_{t}+18\, x^{2}_{t} \,ln\, x_{t}} {12\,(-1+x_t)^4}\,\, ,
\label{F2SM}
\end{eqnarray}
and 
$F^{Beyond}_2\,(y_t, y_t^{\prime})$ is the contribution coming from the charged 
Higgs bosons in $3HDM(O)_2$:
\begin{eqnarray}
F^{Beyond}_2\,(y_t, y_t^{\prime})&=&\frac{1}{m_{t}^2} \,
(\bar{\xi}^{* U}_{N,tt}+\bar{\xi}^{* U}_{N,tc}
\frac{V_{cs}^{*}}{V_{ts}^{*}}) \, (\bar{\xi}^{U}_{N,tt}+\bar{\xi}^{U}_{N,tc}
\frac{V_{cb}}{V_{tb}}) G_{1}(y_t)\nonumber  \, \,  \\
&+&\frac{1}{m_t m_b} \, (\bar{\xi}^{* U}_{N,tt}+\bar{\xi}^{* U}_{N,tc}
\frac{V_{cs}^{*}}{V_{ts}^{*}}) \, (\bar{\xi}^{D}_{N,bb}+\bar{\xi}^{D}_{N,sb}
\frac{V_{ts}}{V_{tb}}) G_{2}(y_t)\nonumber  \, \,  \\ &+&
\frac{1}{m_{t}^2} \,(\bar{\rho}^{* U}_{N,tt}+\bar{\rho}^{* U}_{N,tc}
\frac{V_{cs}^{*}}{V_{ts}^{*}}) \, (\bar{\rho}^{U}_{N,tt}+\bar{\rho}^{U}_{N,tc}
\frac{V_{cb}}{V_{tb}}) G_{1}(y_t')\nonumber  \, \,  \\
&+&\frac{1}{m_t m_b} \, (\bar{\rho}^{* U}_{N,tt}+\bar{\rho}^{* U}_{N,tc}
\frac{V_{cs}^{*}}{V_{ts}^{*}}) \, (\bar{\rho}^{D}_{N,bb}+\bar{\rho}^{D}_{N,sb}
\frac{V_{ts}}{V_{tb}}) G_{2}(y_t')
\label{Fbeyond1}
\end{eqnarray}
with 
\begin{eqnarray}
G_1\,(y)&=& \frac{y}{12\, (-1+y)^4} 
\big( (-1+y)\,(-2-5 \,y+y^2)+6\, y\,ln\,y) \big )\nonumber 
\,\, ,\\ 
G_2\,(y)&=& \frac{1}{2\, (-1+y)^4} \,( y\,(3-4\, y+y^2) + 
2 \,(-1+y)\, y\, ln\,y) \,\, .
\label{G1G2}
\end{eqnarray}
where $y_t=\frac{m_t^2}{m_{H^\pm}}$ and  $y_t'=\frac{m_t^2}{m_{F^\pm}}$
(see appendix B). 
In eq. (\ref{Fbeyond1}) we used the redefinition
\begin{eqnarray}
\xi\,(\rho)^{U,D}=\sqrt{\frac{4 G_{F}}{\sqrt{2}}} \,\, 
\bar{\xi}(\bar{\rho})^{U,D}\,\, .
\label{ksidefn}
\end{eqnarray}

In eq. (\ref{Tfunctot1}), $T_{\mu\nu}^{light}$ contains two different 
parts related to the light internal quarks. The first one is obtained from 
$T^{heavy}_{\mu\nu}$ with the replacement 
$F_2^{3HDM} \rightarrow -F_2^{SM}(0)$ and 
the second one $T^{light}_{2\,\mu\nu}$ is the contribution due to the 
non-vanishing $k_{external}^2/m_{light}^2$ terms 
\begin{eqnarray}
T^{light}_{2\,\mu\nu}&=&-\lambda_t\, \{(\Delta\,i_2-\Delta\, F_1) 
(\not\!{k_1}-\not\!{k_2})\, g_{\mu\nu}\, L+ \Delta\, i_5\, i\, 
\epsilon_{\alpha\mu\nu\beta} \gamma^{\beta} 
(k_1^{\alpha}-k_2^{\alpha})\,L\nonumber \\ &-&2 \Delta\,F_1\, 
(\gamma_{\nu}\,k_{2\,\mu}-\gamma_{\mu}\,k_{1\,\nu})\,L \}
\label{Tlight2}
\end{eqnarray}
where 
\begin{eqnarray}
\Delta F_{1}&=&-{\frac{2}{9}}-{\frac{4}{3}} \frac{Q_{0}(z)}{z}-{\frac{2}{3}} 
Q_{0}(z)\nonumber \,\, , \\
\Delta i_{2}&=&-{\frac{5}{9}}-2 \frac{Q_{-}(z)}{z}+{\frac{8}{3}}
\frac{Q_{0}(z)}{z}-{\frac{2}{3}}Q_{0}(z) \nonumber\,\, ,\\
\Delta i_{5}&=&-1-2  \frac{Q_{-}(z)}{z} \,\, ,
\label{F1i2i5}
\end{eqnarray}
with
\begin{eqnarray}
Q_{0}(z)&=&-2-(u_{+}-u_{-})(ln \frac {u_{-}}{u_{+}}+i\pi)\nonumber\,\, , \\
Q_{-}(z)&=&\frac{1}{2}(ln \frac {u_{-}}{u_{+}}+i\pi)^2\,\, . 
\label{Q0mp} 
\end{eqnarray}
Here the parameter $u_{\pm}$ is  
\begin{eqnarray}
u_{\pm}=\frac{1}{2}(1\pm\sqrt {1-\frac{4}{z}}) \,\, ,
\label{upm}      
\end{eqnarray}
and
\begin{eqnarray}
z=\frac{q^2}{m_i^2},\,\,\, i=u,c \,\, .
\label{zz}
\end{eqnarray}
Finally the function $T_{\mu\nu}$ reads as
\begin{eqnarray}
T_{\mu\nu}=\alpha_R \,
(T^{heavy}_{\mu\nu}+T^{heavy}_{\mu\nu}(F_2^{3HDM}\rightarrow -
F_2^{SM}(0)))+\alpha_L \, T^{light}_{2\,\mu\nu}\,\, .
\label{Tfunctot2}
\end{eqnarray}
Here $\alpha_R$, $\alpha_L$ are real parameters to seperate the parts with
factors $R$ and $L$ in the fuction $T_{\mu\nu}$. With this parametrization  
$\Gamma$ can be written as 
\begin{eqnarray}
\Gamma=\alpha_R^2 \Gamma^R + \alpha_L^2 \Gamma^L + \alpha_R \alpha_L 
\Gamma^{LR}|_{\alpha_L\rightarrow 1, \alpha_R\rightarrow 1} \, .
\label{GammaLR}
\end{eqnarray}

Note that the expressions for the model III case can be obtained by 
disregarding the Yukawa couplings $\bar{\rho}^{U,(D)}_{N,ij}$ in eq. 
(\ref{Fbeyond1}).
\section{$3HDM(O_2)$}
In the multi-Higgs doublet ($n>2$) models, the Higgs sector is extended and 
therefore the number of free parameters, namely, masses of charged and 
neutral Higgs particles, Yukawa couplings, extremely increases. In our
problem we choose $n=3$ and to overcome the difficulty coming from the 
large number of free parameters we consider $3$ Higgs scalars as orthogonal 
vectors in the Higgs flavor space, denoting by the index $m$", 
where $m=1,2,3$. At this stage we introduce a new global $O(2)$ symmetry on 
the Higgs sector \cite{eril3HDM}
\begin{eqnarray}
\phi'_{1}&=&\phi_{1}\nonumber \,\,,\\
\phi'_{2}&=&cos\,\alpha\,\, \phi_2+sin\,\alpha\,\, \phi_3 \,\, , \nonumber \\
\phi'_{3}&=&-sin\,\alpha\,\, \phi_2 + cos\,\alpha\,\, \phi_3\,\,,
\label{trans}
\end{eqnarray}
where $\alpha$ is the global parameter, which represents a rotation of 
the vectors $\phi_2$ and $\phi_3$ along the axis that $\phi_1$ lies, 
in the Higgs flavor space. 
This transformation keeps the kinetic term of 3HDM Lagrangian invariant:
\begin{eqnarray}
{\cal L}_{Kinetic}&=&(D_{\mu} \phi_i)^+ D^{\mu} \phi_i = \nonumber \\
& & (\partial_{\mu} \phi_i^+ + i\frac{g'}{2} B_{\mu} \phi^{+}_i+
i \frac{g}{2} \phi^+_{i} \frac{\vec{\tau}}{2} \vec{W}_{\mu}) 
\nonumber \\
& & (\partial^{\mu} \phi_i - i\frac{g'}{2} B^{\mu} \phi_i-
i \frac{g}{2} \phi_{i} \frac{\vec{\tau}}{2} \vec{W}^{\mu}) 
\label{kinetic}
\end{eqnarray}
where 
\begin{eqnarray}
\phi_{i}=\left(\begin{array}{c c} 
\phi^{+}\\ \phi^0\end{array}\right)  \,\, i=1,2,3\, .
\label{phi}
\end{eqnarray}
The invariance of the potential term 
\begin{eqnarray}
V(\phi_1, \phi_2,\phi_3 )&=&c_1 (\phi_1^+ \phi_1-v^2/2)^2+
c_2 (\phi_2^+ \phi_2)^2 \nonumber \\ &+& 
c_3 (\phi_3^+ \phi_3)^2+
c_4 [(\phi_1^+ \phi_1-v^2/2)+ \phi_2^+ \phi_2+\phi_3^+ \phi_3]^2
\nonumber \\ &+& 
c_5 [(\phi_1^+ \phi_1) (\phi_2^+ \phi_2)-(\phi_1^+ \phi_2)(\phi_2^+ \phi_1)]
\nonumber \\ &+& 
c_6 [(\phi_1^+ \phi_1) (\phi_3^+ \phi_3)-(\phi_1^+ \phi_3)(\phi_3^+ \phi_1)]
\nonumber \\ &+& 
c_7 [(\phi_2^+ \phi_2) (\phi_3^+ \phi_3)-(\phi_2^+ \phi_3)(\phi_3^+ \phi_2)]
\nonumber \\ &+& 
c_8 [Re(\phi_1^+ \phi_2)]^2 +c_9 [Re(\phi_1^+ \phi_3)]^2 +
c_{10} [Re(\phi_2^+ \phi_3)]^2 \nonumber \\ &+&
c_{11} [Im(\phi_1^+ \phi_2)]^2 +c_{12} [Im(\phi_1^+ \phi_3)]^2 +
c_{13} [Im(\phi_2^+ \phi_3)]^2 +c_{14}
\label{potential}
\end{eqnarray}
can be obtained if the following conditions on the free parameters 
are satisfied:
\begin{eqnarray}
& & c_5=c_6\,\,\, , c_8=c_9\,\, , c_{11}=c_{12}\,\, , \nonumber \\
& & c_2=c_3=c_7=c_{10}=0 \,\, .
\label{ci}
\end{eqnarray}
This implies that the masses of new particles are the same as the older
ones, namely, 
\begin{eqnarray}
m_{F^\pm}=m_{H^\pm}=c_5 \frac{v^2}{2}\nonumber \,\, , \\
m_{H^3}=m_{H^1}=c_{8} \frac{v^2}{2}\nonumber \,\, , \\
m_{H^4}=m_{H^2}=c_{11} \frac{v^2}{2}\nonumber \,\, , \\
\label{mass2}
\end{eqnarray}
Further, the application of this transformation to the Yukawa Lagrangian 
(eq.(\ref{lagrangian})) keeps it invariant if the transformed 
Yukawa matrices satisfy the expressions
\begin{eqnarray}
\bar{\xi}^{\prime U(D)}_{ij}&=& \bar{\xi}^{U (D)}_{ij} cos\, \alpha+
\bar{\rho}^{U(D)}_{ij} sin\, \alpha\,\, \nonumber ,\\
\bar{\rho}^{\prime U (D)}_{ij}&=&-\bar{\xi}^{U (D)}_{ij} sin\, \alpha+
\bar{\rho}^{U (D)}_{ij} cos\, \alpha \,\, .
\label{yuktr}
\end{eqnarray}
and therefore 
\begin{eqnarray}
(\bar{\xi}^{\prime U(D)})^ 
+ \bar{\xi}^{\prime U (D) } +
(\bar{\rho}^{\prime U (D)})^+\bar{\rho}^{\prime U (D) }=
(\bar{\xi}^{U (D)})^+\bar{\xi}^{U (D)} +
(\bar{\rho}^{U (D)})^+ \bar{\rho}^{U (D) }\,\, , 
\label{yukinv}
\end{eqnarray}
which allows us the following possible parametrization of the Yukawa matrices 
$\bar{\xi}^{U(D)}$ and $\bar{\rho}^{U(D)}$: 
\begin{eqnarray}
\bar{\xi}^{U (D)}=\epsilon^{U(D)} cos\,\theta \nonumber \,\, ,\\
\bar{\rho}^{U}=\epsilon^{U} sin\,\theta \nonumber \,\, ,\\ 
\bar{\rho}^{D}=i \epsilon^{D} sin\,\theta \,\, ,
\label{yukpar}
\end{eqnarray}
where $\epsilon^{U(D)}$ are real matrices satisfy the equation 
\begin{eqnarray}
(\bar{\xi}^{\prime U(D)})^+ \bar{\xi}^{\prime U (D) } +
(\bar{\rho}^{\prime U (D)})^+\bar{\rho}^{\prime U (D) }=
(\epsilon^{U(D)})^T \epsilon^{U(D)} 
\label{yukpareq}
\end{eqnarray}
Here $T$ denotes transpose operation. In eq. (\ref{yukpar}),  we take 
$\bar{\rho}^{D}$ complex to carry all $CP$ violating effects in the third 
Higgs scalar. 

Therefore we can reduce the number of free parameters taking the new
charged and neutral boson masses as the same as the older ones and 
connecting the Yukawa matrices $\bar{\xi}^{U(D)}$ and 
$\bar{\rho}^{U(D)}$ using the expression eq. (\ref{yukpareq}).

Note that, neglecting the off-dioganal Yukawa couplings, the expression for 
$F^{Beyond}_2\,(y_t, y_t^{\prime})$ (eq. (\ref{Fbeyond1})) can be written as  
\begin{eqnarray}
F^{Beyond}_2\,(y_t)&=&\frac{1}{m_{t}^2} \,
(\bar{\epsilon}^{U}_{N,tt})^2 \, G_{1}(y_t) +
\frac{1}{m_t m_b} \, \bar{\epsilon}^{U}_{N,tt}\, \bar{\epsilon}^{D}_{N,bb}
G_{2}(y_t)\, (cos^2\,\theta + i\, sin^2\,\theta) 
\label{Fbeyond2}
\end{eqnarray}
\end{appendix}
\newpage 
\newpage
\end{document}